\documentclass[doublecol]{epl2} 
\usepackage{amsmath}
\usepackage{amssymb}
\usepackage{stfloats}
\usepackage[capitalise]{cleveref}
\renewcommand{\Re}{\operatorname{Re}}
\renewcommand{\Im}{\operatorname{Im}}

\title{Properties of Non-topological Solitons in Two-dimensional Model With Resurrected Conformal Symmetry}
\shorttitle{Non-topological Solitons in CFT} 

\author{Yu. Galushkina\inst{1} \and E. Kim\inst{1,2} \and E. Nugaev\inst{1} \and Ya. Shnir\inst{3}}
\shortauthor{Yu. Galushkina \etal}

\institute{                    
  \inst{1} Institute for Nuclear Research of RAS- prospekt 60-letiya Oktyabrya 7a, Moscow, 117312\\
  \inst{2} Moscow Institute of Physics and Technology- Institutsky lane 9, Dolgoprudny, Moscow region, 141700\\
  \inst{3} Joint Institute for Nuclear Research- Joliot-Curie St 6, Dubna, Moscow region, 141980\\
}

\abstract{
We study properties of non-topological solitons in two-dimensional conformal field theory. The spectrum of linear perturbations on these solutions is found to be trivial, containing only symmetry-related zero modes. The interpretation of this feature is given by considering the relativistic generalization of our theory in which the conformal symmetry is violated. It is explicitly seen that the restoration of this symmetry leads to the absence of decay/vibrational modes.}

\begin{document}

\maketitle

\section{Introduction} 


Symmetries play a special role in modern physics. Among many known physical symmetries, one stands out notably: the conformal symmetry. It was previously established that conformal symmetry is related to scale invariance (dilatations), and thus the theory lacks characteristic length scales. In this case, dynamical symmetries of space and time emerge as a symmetry group that is more complex than the Poincar\'{e} group, which is common for particle physics. In the non-relativistic limit something similar happens to the Galilei group as it is completed to the Schr\"{o}dinger group for a free theory \cite{Bargmann:1954gh,PhysRevD.5.377,Niederer:1972zz} (a recent review is given in \cite{Duval:2024eod}).   


Although interactions typically violate symmetries, there are some special cases in which conformal symmetry is preserved in the non-linear model \cite{deKok:2007ur}. Preservation of an exact scale invariance and conformal symmetry is provided by choosing the specific power of non-linearity in Lagrangian depending on the dimensionality of space.   


Non-linear equations of motion may provide soliton solutions that propagate without dispersion and maintain their shape. One of the most elegant examples that can be observed in Nature is the non-topological bright soliton \cite{shabat1972exact}. These objects are often used for the modeling of lumps of Bose condensate. Recently, this approach has been used in various phenomenological models for the description of ultralight dark matter \cite{Hu:2000ke}. In this scope, the dynamics of solitons is of great interest as their excitations provide valuable insights into stability of solutions.    



In this paper, we study non-relativistic bright solitons and their properties within the framework of two-dimensional conformal field theory. For our model, we have found a continuous branch of analytical solutions of different sizes that represent bright solitons. Remarkably, their properties, i.e., energy, $U(1)$ charge and linear stability, do not depend on the width. In order to examine the role of conformal symmetry in these processes, we study the violation of this symmetry in the relativistic theory. In the generalized model, we have found a decay mode that is in agreement with the Vakhitov-Kolokolov criterion \cite{Vakhitov:1973lcn}.   


Firstly, we introduce a two-dimensional interacting model that preserves dilatations and conformal symmetry. A brief overview of the symmetries of our model is also provided. Then, we discuss an explicit solution for conformal bright solitons and describe their properties. We also consider the aspects of relativistic generalization and conformal symmetry restoration.

\section{The model and symmetries}

In this section, we introduce non-relativistic Lagrangian\footnote[1]{Bearing in mind the following relativistic generalization, we use the natural system of units $\hbar=c=1$.} of complex scalar field
\begin{equation} \label{NR lagrangian}
    \mathcal{L}_{GP6} = i\psi^{\ast}\dot{\psi} - \frac{1}{2m}\nabla\psi^{\ast}\nabla\psi + \frac{\lambda}{24m^{3}}\left(\psi^{\ast}\psi \right)^{3}
\end{equation}
of our model in one space dimension. The corresponding equation of motion is a quintic non-linear Schr\"{o}dinger equation (NLSE), which is also known as the Gross-Pitaevskii equation with the quintic term   
\begin{equation}\label{NLSE}
    i\frac{\partial}{\partial t}\psi = \left[-\frac{\nabla^{2}}{2m} - \frac{\lambda}{8m^{3}}\left( \psi^{\ast}\psi \right)^{2} \right]\psi.
\end{equation}

\begin{table*}[tp]
\begin{center}
\caption{Transformations and infinitesimal generators of Schr\"{o}dinger group.}
\label{tab.1}
\setlength{\tabcolsep}{15pt}
\renewcommand{\arraystretch}{1.7}
\begin{tabular}{ |p{4cm}|p{4cm}|p{4cm}|  } 
 \hline
 \multicolumn{3}{|c|}{Schr\"{o}dinger group} \\
 \hline
 Subgroup & Transformations & Infinitesimal generators $G$  \\
 \hline
 Time translation   & $t^{'}=t+\beta $ & $\frac{\partial}{\partial t}$\\
 \hline
 Space translation & $x^{'}=x+a$ & $\frac{\partial}{\partial x}$ \\
\hline
 Rotation & $x^{'}=x$ & $1$\\
 \hline
 Galilean boost & $x^{'}=x + v\cdot t$ & $t\frac{\partial}{\partial x}-imx$\\
\hline
 Dilatation &   $t^{'}=e^{2\sigma}t$, $x^{'}=e^{\sigma}x$  & $2t\frac{\partial}{\partial t}+x\frac{\partial}{\partial x}+\frac{1}{2}$ \\
\hline
 Special conformal symmetry & $t^{'}=\frac{t}{1+\eta t}$, $x^{'}=\frac{x}{1+\eta t}$ & $\frac{i m x^{2}}{2}-\frac{t}{2}-xt\frac{\partial}{\partial x}-t^{2}\frac{\partial}{\partial t}$ \\
 \hline
\end{tabular}
\end{center}
\end{table*}

Physically, the quintic term may simulate three-body interactions and/or deviations of the dimensionality of the condensate  (see  e.g.\cite{gammal2000atomic,Avelar:2008va,kengne2008bose}).

Both Eq.(\ref{NLSE}) and the Lagrangian (\ref{NR lagrangian}) possess a global $U(1)$ symmetry along with a Schr\"{o}dinger group invariance (see \cite{deKok:2007ur} and references therein). An internal global $U(1)$ symmetry is responsible for the conservation of a particle number
\begin{equation}\label{particle number N}
    N = \int_{-\infty}^{\infty} dx \psi^{\ast}(t,x)\psi(t,x).
\end{equation}

In addition to that, Schr\"{o}dinger group is a much larger and more fruitful group of space and time transformations than the Poincar\'{e} group. Among many interesting features of this group, we especially highlight the emergence of a projective representation of the Galilei group \cite{Bargmann:1954gh}. 
We provide a brief description of this group in Table \ref{tab.1}. The preservation of the Schr\"{o}dinger group is ensured by our choice of the potential term of the model (\ref{NR lagrangian}). In ref. \cite{deKok:2007ur}, it was established that unbroken dilatation and conformal symmetry are present in a model with potential term $|\psi|^{2n}$ that satisfies relation
\begin{equation}
    nd=d+2,
\end{equation}
where d is the number of space dimensions. Note that the common cubic NLSE that possesses bright soliton solutions leads to the Schr\"{o}dinger group breaking to the Galilei group. In order to maintain bright solitons in a conformal field theory, we need to consider the coupling $\lambda$ in Lagrangian (\ref{NR lagrangian}) to be positive-defined, so that the potential is of an attractive kind. 


\section{Soliton and excitations}


Now, we are ready to provide a bright soliton solution of Eq.(\ref{NLSE}) using following ansatz $\psi(t,x)=e^{i\mu t}f(x)$. The equation of motion can be solved analytically, and the soliton has a form of
\begin{equation}\label{NR soliton}
    \psi(t,x)=e^{i\mu t} \left(\frac{24 m^3 \mu}{\lambda }\right)^{\frac{1}{4}} \sqrt{\text{sech}\left(\sqrt{8 m \mu} \cdot x\right)},
\end{equation}
where $\mu$ is a continuous dimensionful parameter. 

It is worth studying the integral characteristics of these solutions, such as the $U(1)$ charge and the energy functional. Thus, straightforward calculations show that
\begin{equation}\label{NR integral}
    \begin{split}
        & N = \int_{-\infty}^{\infty} dx |\psi(t,x)|^{2} = \frac{\sqrt{3} \pi m }{\sqrt{\lambda}}, \\
        & H = \int_{-\infty}^{\infty} dx \left[\frac{1}{2m}|\nabla \psi(t,x)|^{2} - \frac{\lambda}{24m^{3}}|\psi(t,x)|^{6} \right] = 0.
    \end{split}
\end{equation}
At first sight, this result seems odd: how come none of the characteristics above depend on parameter $\mu$? The answer is in the presence of dilatation and conformal symmetries in Table \ref{tab.1}. To show this explicitly, we redefine $e^{\sigma}$ as $\sqrt{2m\mu}$, so that $t^{'}=2m\mu t$ and $x^{'}=\sqrt{2m\mu}\cdot x$ for dilatation transformations. The complex field $\psi$ transforms as $\psi^{'} = (2m\mu)^{-\frac{1}{4}}\psi$. In new terms, NLSE can be rewritten in a form that is $\mu$-independent
\begin{equation}
    \nabla_{x^{'}}^{2}\psi^{'} = \psi^{'} - \frac{\lambda}{4m^{2}}\left\vert\psi^{'}\right\vert^{4}\psi^{'}.
\end{equation}
This allows us to expand the integral characteristics as
\begin{equation}
\begin{split}
    & N = \frac{\sqrt{2m\mu}}{\sqrt{2m\mu}}\int_{-\infty}^{\infty} dx^{'} \left\vert\psi^{'}\left(t^{'},x^{'}\right)\right\vert^{2}, \\
    & H = \frac{(2m\mu)^{\frac{3}{2}}}{\sqrt{2m\mu}}\int_{-\infty}^{\infty} dx^{'} \left[\frac{1}{2m}\left\vert\nabla_{x^{'}} \psi^{'}\left(t^{'},x^{'}\right)\right\vert^{2} - \right.\\
    &\left. -\frac{\lambda}{24m^{3}}\left\vert\psi^{'}\left(t^{'},x^{'}\right)\right\vert^{6} \right]=\frac{(2m\mu)^{\frac{3}{2}}}{\sqrt{2m\mu}}\cdot 0,
\end{split}
\end{equation}
where the latter is a direct consequence of conformal symmetry \cite{Ghosh:2001an}.

Relations (\ref{NR integral}) impose a constraint on both the energy and the $U(1)$ charge of bright solitons (\ref{NR soliton}). In this case, the study of the spectrum of linear perturbations of the soliton is of great interest. Using a perturbation ansatz
\begin{equation}
    \psi_{p}(t,x) =\psi(t,x) + \delta\psi(t,x) =e^{i\mu t}f(x) + \delta\psi(t,x)
\end{equation}
one can derive the linearized equation of motion

\begin{equation}\label{NR linear. eq.}
\begin{split}
    &i\frac{\partial}{\partial t}\delta \psi(t,x) = -\frac{\nabla^{2}}{2m}\delta \psi(t,x) - \frac{\lambda}{8m^{3}}|\psi(t,x)|^{4}\times\\
    &\times \left(3\cdot\delta\psi(t,x)+2\cdot\delta\psi^{\ast}(t,x)e^{2i\mu t} \right).
\end{split}
\end{equation}
Prior to studying any decay/oscillating modes we consider symmetry-related zero modes as they have a simple form
\begin{equation}
    \delta\psi_{0}(t,x) = G\psi(t,x),
\end{equation}
where $G$ is an infinitesimal symmetry generator from Table \ref{tab.1} or a generator of $U(1)$ symmetry $G=i$. 

The general ansatz for linear perturbations of the complex field $\psi$ can be written as \cite{1970JMP....11.1336A}
\begin{equation}
\delta\psi(t,x) = e^{i\mu t}\left( e^{i\gamma t}\eta(t,x)+e^{-i\gamma^{\ast}t}\xi^{\ast}(t,x) \right),
\end{equation}
where two distinct cases should be highlighted. By setting the parameter $\gamma$ and the functions $\eta$, $\xi$ to be real we study the vibrational modes of bright soliton. Considering decay modes requires redefinition $\gamma\rightarrow -i\gamma, \gamma\in \mathbb{R}$ while the functions $\eta,\xi\in\mathbb{C}$. Dilatation symmetry along with an additional scaling of the field $\psi$ by factor $\frac{1}{\lambda^{\frac{1}{4}}}$ allows us to write simplified linearized equations of motion\footnote[2]{For a moment we neglect the prime sign in the next two equations, however it is crucial to remember that time and space coordinates are transformed as $t^{'}=2m\mu t$, $x^{'}=\sqrt{2m\mu}\cdot x$ and $\psi^{'} = \left(\frac{2m\mu}{\lambda}\right)^{-\frac{1}{4}}\psi$.} for vibrational modes
\begin{equation}
    \begin{split}
        & \nabla^{2}\eta = \left( 1+\frac{\gamma_{osc.}}{\mu} \right)\eta - \frac{1}{4m^{2}}f^{4}(3\eta+2\xi), \\
        & \nabla^{2}\xi = \left( 1-\frac{\gamma_{osc.}}{\mu} \right)\xi - \frac{1}{4m^{2}}f^{4}(3\xi+2\eta).
    \end{split}
\end{equation}
and for decay modes, we redefine $\xi\equiv(\eta+\xi^{\ast})$
\begin{equation}
    \begin{split}
        & \nabla^{2}\Re\xi = \Re\xi + \frac{\gamma_{dec.}}{\mu}\Im\xi - \frac{5}{4m^{2}}f^{4}\Re\xi , \\
        & \nabla^{2}\Im\xi = \Im\xi - \frac{\gamma_{dec.}}{\mu}\Re\xi - \frac{1}{4m^{2}}f^{4}\Im\xi .
    \end{split}
\end{equation}

An extensive numerical scanning (we applied this approach in \cite{Galushkina:2024iad}) of normalizable modes which are localized in spatial dimension has failed to find any modes at any value of the parameter $\mu$ other than zero modes. This result is in agreement with the well-known Vakhitov-Kolokolov criterion, since the bright soliton (\ref{NR soliton}) does not belong to the stability region
\begin{equation}
    \frac{\mu}{N}\frac{d}{d\mu}N<0
\end{equation}
or the instability region

\begin{equation}
\frac{\mu}{N}\frac{d}{d\mu}N>0.
\end{equation}

In the next section, we argue that there are indeed no vibrational/decay modes in the conformal theory (\ref{NR lagrangian}) as can be seen from a straightforward relativistic generalization.





\section{Relativistic generalization}

In order to provide relativistic generalization of the model (\ref{NR lagrangian}) we use a simple relation between the relativistic field $\phi$ and the non-relativistic field $\psi$ that has the form 
\begin{equation}\label{NR ansatz}
    \phi(t,x) = \frac{1}{\sqrt{2m}}e^{-imt}\psi(t,x).
\end{equation}
Using relation (\ref{NR ansatz}), we are able to write down the following Lorentz-invariant Lagrangian
\begin{equation}\label{Rel. model}
    \mathcal{L} = \partial_{\mu}\phi^{\ast}\partial^{\mu}\phi - m^{2}\phi^{\ast}\phi + \frac{\lambda}{3}\left( \phi^{\ast}\phi \right)^{3}.
\end{equation}

\begin{figure}[htb]
    \centering
    \includegraphics[width=1\linewidth]{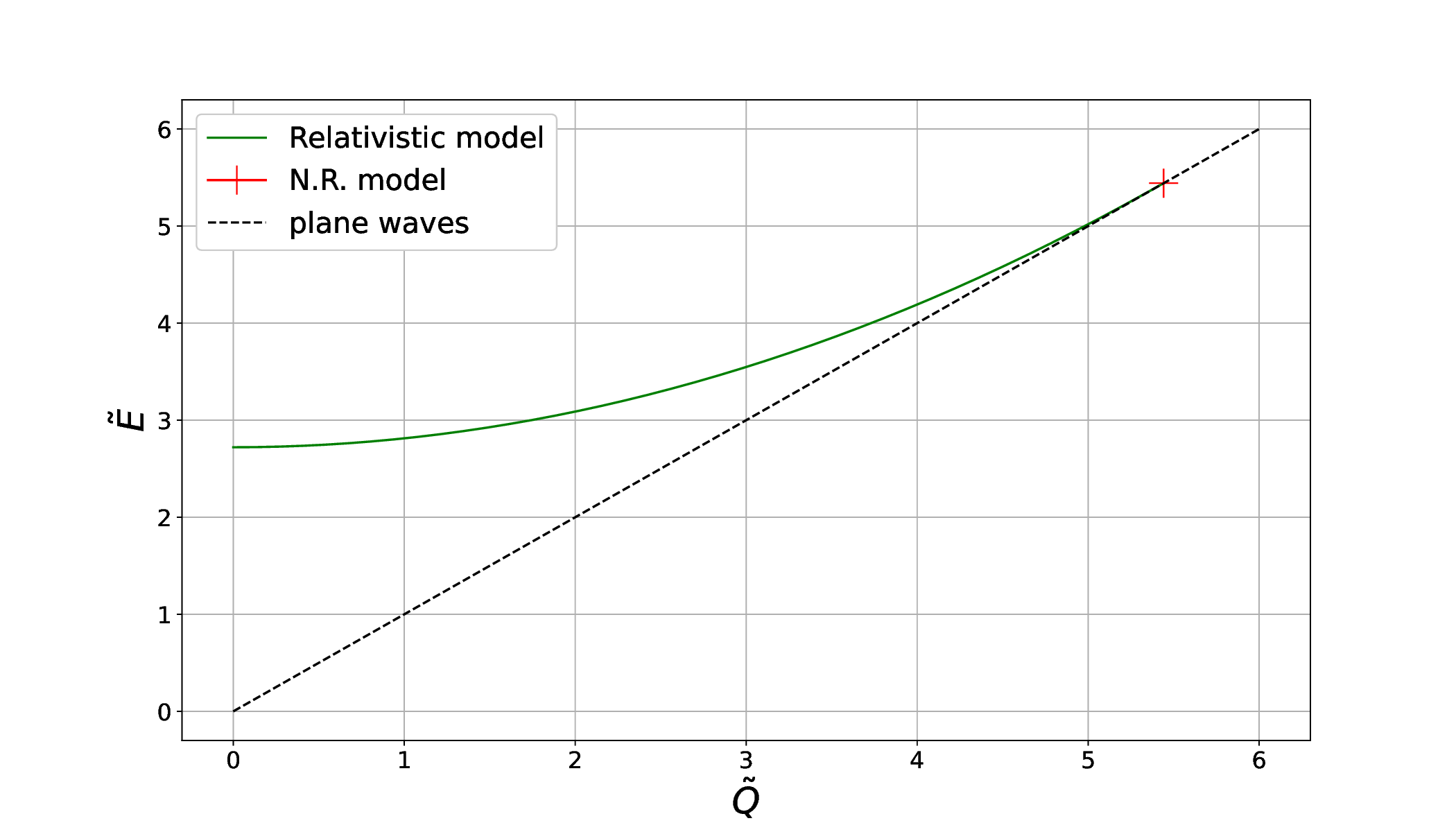}
    \caption{Dimensionless integral characteristics $\tilde{E}=\frac{E}{m}, \tilde{Q}=Q$ of relativistic model (\ref{Rel. model}) calculated at parameter $\frac{\lambda}{m^{2}}=1$. Dashed line represents plane waves solutions of mass $mQ$ and the cross mark corresponds to conformal theory (\ref{NR lagrangian}).}
    \label{fig1}
\end{figure}

This theory also supports a soliton solution that can be written as
\begin{equation}\label{Rel. soliton}
\begin{split}
    &\phi(t,x)=e^{-i\omega t}g_{\omega}(x)=e^{-i\omega t}\left(\frac{3 \left(m^2-\omega^2\right)}{\lambda }\right)^{\frac{1}{4}}\times\\
    &\times \sqrt{\text{sech}\left(2\sqrt{m^2-\omega^2}\cdot x\right)}.
    \end{split}
\end{equation}
In fact, the bright soliton (\ref{NR soliton}) can be derived from the relativistic soliton by defining $\mu=m-\omega$ and keeping only linear order in $\mu$ and using the relation (\ref{NR ansatz}). Once again, it is helpful to calculate the integral characteristics (see Fig.\ref{fig1}). The $U(1)$ charge is
\begin{equation}\label{Q}
    Q = 2\omega\int_{-\infty}^{\infty} dx \left\vert\phi(t,x) \right\vert^{2} = \frac{\sqrt{3}\pi \omega}{\sqrt{\lambda}}
\end{equation}
and the energy functional results in
\begin{equation}\label{E}
\begin{split}
    &E = \int_{-\infty}^{\infty}dx\left[ \left\vert \dot{\phi} \right\vert^{2} + \left\vert \nabla\phi \right\vert^{2} + m^{2}\left\vert \phi \right\vert^{2}- \frac{\lambda}{3}\left\vert \phi \right\vert^{6} \right] = \\
    & = \frac{\sqrt{3}\pi (m^{2}+\omega^{2})}{2\sqrt{\lambda}}.
    \end{split}
\end{equation}
It can be directly checked that the differential relation $\frac{dE}{dQ}=\omega$ is satisfied. 

We examine our relativistic generalization by comparing the integral characteristics (\ref{NR integral}) and (\ref{Q},\ref{E}) in the limit $\omega\to m$   
\begin{equation}
    \begin{split}
        & Q\vert_{\omega\to m} = \frac{\sqrt{3}\pi m}{\sqrt{\lambda}}=N, \\
        & \left[E-\omega Q\right]\vert_{\omega\to m} = 0 = H.
    \end{split}
\end{equation}

The differential relation $\frac{dE}{dQ}=\omega$ and Fig.(\ref{fig1}) show that the Vakhitov-Kolokolov instability criterion $\frac{dQ}{d\omega}>0$ is fulfilled and we should be able to find decay modes on the soliton (\ref{Rel. soliton}). Following scaling
\begin{equation}
    \begin{split}
        & \tilde{x} = x\sqrt{m^2-\omega^{2}}; \\
        & \tilde{g} = \frac{g_{\omega}\lambda^{\frac{1}{4}}}{(m^{2}-\omega^{2})^{\frac{1}{4}}},
    \end{split}
\end{equation}
allows us to write linearized equations of motion for decay modes $\delta\phi(t,x) = e^{-i\omega t}e^{\gamma_{dec.}t}\left(\Re\xi(x)+i\Im\xi(x)\right)$
\begin{equation}\label{Lin. eqs decay modes}
    \begin{split}
        & \tilde{\nabla}^{2}\Re\xi = \frac{\left(m^{2}-\omega^{2}+\gamma_{dec.}^{2}\right)\Re\xi+2\omega\gamma_{dec.}\Im\xi}{m^{2}-\omega^{2}} - \\
        & - 5\tilde{g}^{4}\Re\xi , \\
        & \tilde{\nabla}^{2}\Im\xi = \frac{\left(m^{2}-\omega^{2}+\gamma_{dec.}^{2}\right)\Im\xi-2\omega\gamma_{dec.}\Re\xi}{m^{2}-\omega^{2}} - \\
        & - \tilde{g}^{4}\Im\xi .
    \end{split}
\end{equation}

\begin{figure}[htb]
    \centering
    \includegraphics[width=1\linewidth]{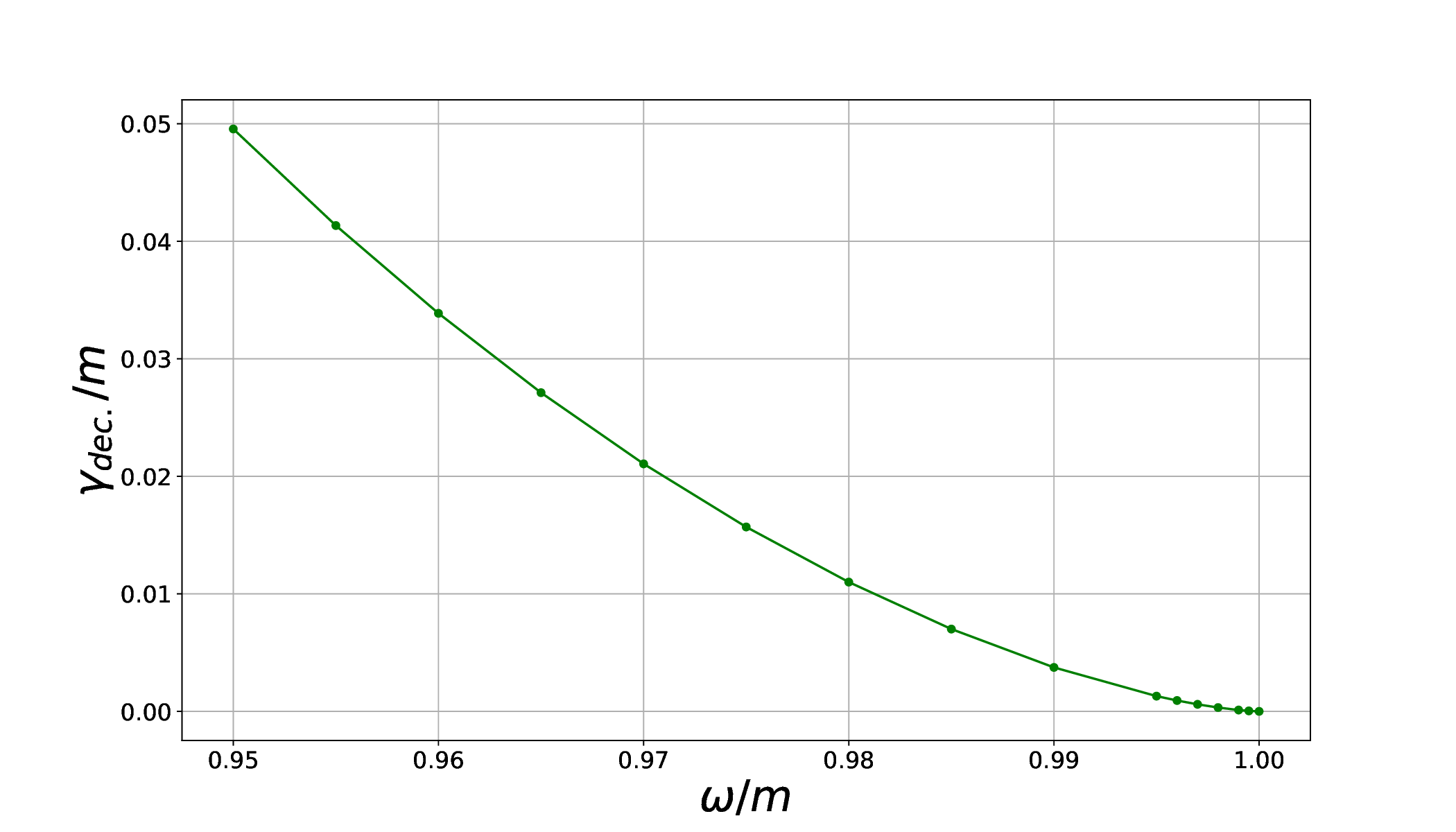}
    \caption{Spectrum of decay modes that are described by Eqs.(\ref{Lin. eqs decay modes}). In the limit $\omega\to m$ parameter $\gamma_{dec.}$ tends to zero as $ C\cdot(m-\omega)^{1.506}$.}
    \label{fig2}
\end{figure}

Numerical scanning of the decay modes spectrum in presented in Fig.\ref{fig2}. It can be seen that in the limit $\omega\to m$ parameter $\gamma_{dec.}$ tends to zero. While $\frac{\gamma_{dec.}}{\omega}\ll 1$ decay modes might be generated by expanding a soliton solution in perturbation series as
\begin{equation}\label{Perturbation series}
\begin{split}
    & i\phi_{p}(t,x)=ie^{-i\left(1+i\frac{\gamma}{\omega}\right) \omega t}g_{1+i\frac{\gamma}{\omega}}(x)\approx e^{-i\omega t} \left( 1 + \gamma t \right)\times \\
    & \times \left(ig_{\omega}(x) -\gamma\frac{\partial g_{\omega}(x)}{\partial \omega}  \right).
    \end{split}
\end{equation}

\begin{figure}[htb]
    \centering
    \includegraphics[width=1\linewidth]{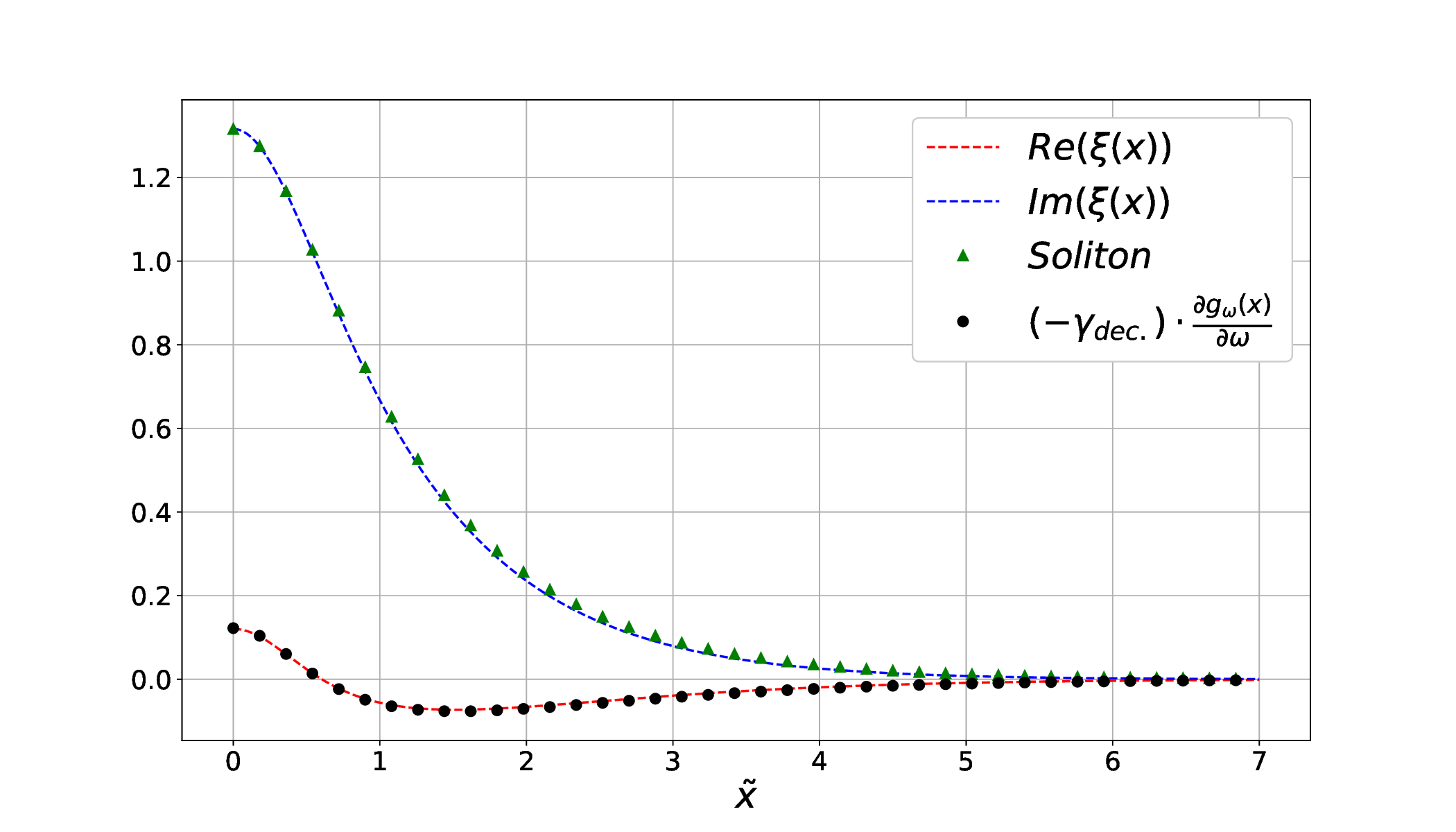}
    \caption{Decay mode profile at $\omega/m=0.99$ and $\gamma_{dec.}/m=3.74\cdot 10^{-3}$. Scaled soliton profile and $\displaystyle{(-\gamma_{dec.})\frac{\partial g_{\omega}(x)}{\partial \omega}}$ are added for comparison.}
    \label{fig3}
\end{figure}

Comparison with the expansion of decay mode ansatz
\begin{equation} \label{rel. decay modes}
    \begin{split}
    & \delta\phi(t,x) = e^{i\omega t}e^{\gamma t}\left( \Re\xi+i\Im\xi \right)\approx \\
    & \approx e^{i\omega t}(1+\gamma t)\left( \Re\xi+i\Im\xi \right)
\end{split}
\end{equation}
helps to evaluate that $\Re\xi=-\gamma\frac{\partial g_{\omega}(x)}{\partial \omega}$ and $\Im\xi=g_{\omega}(x)$. Validating these calculations requires a decay mode profile and we provide it in Fig.\ref{fig3}. A close look at the figure tells us that we have found the mode that is close to the perturbed field (\ref{Perturbation series}). Thus, we can interpret the results of the previous section that bright solitons of conformal field theory (\ref{NR lagrangian}) do not support any non-trivial modes as a consequence of behavior of decay modes (\ref{rel. decay modes}) in the non-relativistic limit ($\omega \to m$). As the parameter $\gamma_{dec.}$ goes to zero $\Re\xi=-\gamma_{dec.}\frac{\partial g_{\omega}(x)}{\partial \omega}$ vanishes and the decay mode transforms into a zero mode related to $U(1)$ symmetry.

\section{Outlook}

In this paper, we have considered Lorentz-invariant field theory that supports analytical non-topological solitons. It was shown that the global charge of $U(1)$ symmetry approaches its maximal value in the non-relativistic limit. Contrary to common theories with solitons, this value is not associated with a cusp, but rather is an isolated point. We have studied that in non-relativistic limit conformal symmetry emerges and plays a crucial role. Specifically, the energy and the global charge do not depend on the scale of the soliton solution. This example highlights the importance of relativistic corrections, as they lead to exponential decay modes. Our results were validated analytically and numerically by using perturbation theory.   

One can argue that our relativistic generalization should also include the 2-body interaction term and even more specific modifications (see the example in \cite{Bowcock:2008dn}). 
As a natural continuation of our work, we plan \cite{galushkina2025cftapproachrotatingfieldlumps} to study more physical planar solutions similar to those in \cite{Jackiw:1990tz}. Dilatations and conformal symmetry are preserved at the classical level, although quantum corrections might break these symmetries \cite{deKok:2007ur}. Moreover, it might be fruitful to generalize our results for other representations of the Poincar\'{e} group. This study may be useful for the numerical modeling of ultralight dark matter \cite{Preskill:1982cy,Dine:1982ah,Levkov:2016rkk} and Bose stars, see, e.g. \cite{Jetzer:1991jr,Liebling:2012fv,Suarez:2013iw}.

\acknowledgments
Numerical studies of relativistic decay modes were supported by the grant RSF 22-12-00215-$\Pi$. The work of E. Kim was supported by the Foundation for the Advancement of Theoretical Physics and Mathematics BASIS.

\end{document}